# Resolving the gravitational redshift within a millimeter atomic sample

Tobias Bothwell[1,*], Colin J. Kennedy[1,2], Alexander Aeppli[1], Dhruv Kedar[1], John M. Robinson[1], Eric Oelker[1,3], Alexander Staron[1], and Jun Ye[1,*]

[1]JILA, National Institute of Standards and Technology and University of Colorado

Department of Physics, University of Colorado

Boulder, Colorado 80309-0440, USA

[2]Present Address: Honeywell Quantum Solutions, Broomfield, Colorado 80021

[3]Present Address: Physics Department, University of Glasgow

**Einstein's theory of general relativity states that clocks at different gravitational potentials tick at different rates – an effect known as the gravitational redshift[1]. As fundamental probes of space and time, atomic clocks have long served to test this prediction at distance scales from 30 centimeters to thousands of kilometers[2–4]. Ultimately, clocks will study the union of general relativity and quantum mechanics once they become sensitive to the finite wavefunction of quantum objects oscillating in curved spacetime. Towards this regime, we measure a linear frequency gradient consistent with the gravitational redshift within a single millimeter scale sample of ultracold strontium. Our result is enabled by improving the fractional frequency measurement uncertainty by more than a factor of 10, now reaching $7.6\times10^{-21}$. This heralds a new regime of clock operation necessitating intra-sample corrections for gravitational perturbations.**

Modern atomic clocks embody Arthur Schawlow's motto to "never measure anything but frequency." This deceptively simple principle, fueled by the innovative development of laser science and quantum technologies based on ultracold matter, has led to dramatic progress in clock performance. Recently, clock measurement precision reached the mid-19$^{th}$ digit in one hour[5,6], and three atomic species achieved systematic uncertainties corresponding to an error equivalent to less than 1 s over the lifetime of the Universe[7–10]. Central to this success in neutral atom clocks is the ability to maintain extended quantum coherence times while using large ensembles of atoms[5,6,11]. The pace of progress has yet to slow. Continued improvement in measurement precision and accuracy arising from the confluence of metrology and quantum information science[12–15] promises discoveries in fundamental physics[16–20].

Clocks fundamentally connect space and time, providing exquisite tests of the theory of general relativity. Hafele and Keating took cesium-beam atomic clocks aboard commercial airliners in 1971, observing differences between flight-based and ground-based clocks consistent with special and general relativity[21]. More recently, RIKEN researchers compared two strontium optical lattice clocks (OLCs) separated by 450 m in the Tokyo Skytree, resulting in the most precise terrestrial redshift measurement to date[22]. Proposed satellite-based measurements[23,24] will provide orders of magnitude improvement to current bounds on gravitational redshifts[3,4]. Concurrently, clocks are anticipated to begin playing important roles for relativistic geodesy[25]. In 2010 Chou *et al.*[2] demonstrated the precision of their Al$^+$ clocks by measuring the gravitational redshift resulting from lifting one clock vertically by 30 cm in 40 hours. With a decade of advancements, today's leading clocks are poised to enable geodetic surveys of elevation at the sub-centimeter level on Earth, a result unobtainable with other current techniques[26].

Atomic clocks strive to simultaneously optimize measurement precision and systematic uncertainty. For traditional OLCs operated with one-dimensional (1D) optical lattices, achieving low instability has involved the use of high atom numbers at trap depths sufficiently large to suppress tunnelling between neighboring lattice sites. While impressive performance has been achieved, effects arising from atomic interactions and AC Stark shifts associated with the trapping light challenge advancements in OLCs. Here we report a new operational regime for 1D OLCs, both resolving the gravitational redshift across our atomic sample and synchronously measuring a fractional frequency uncertainty of 7.6×10$^{-21}$ between two uncorrelated regions. Our system employs ~100,000 $^{87}$Sr atoms at ~100 nK loaded into a shallow, large waist optical lattice, reducing both AC Stark and density shifts. Motivated by our earlier work on spin-orbit coupled lattice clocks[27,28], we engineer atomic interactions by operating at a 'magic' trap depth, effectively removing collisional frequency shifts. These advances enable record optical atomic coherence (37 s) and single clock stability (3.1×10$^{-18}$ at 1 s) using macroscopic samples, paving the way toward lifetime limited OLC operation.

Central to our experiment is an in-vacuum optical cavity (Fig. 1a and Methods) for power enhancement of the optical lattice. The cavity (finesse 1300) ensures wavefront homogeneity of our 1D lattice while the large beam waist (260 μm) reduces the atomic density by an order of magnitude compared to our previous system[10]. We begin each experiment by trapping fermionic $^{87}$Sr atoms into the 1D lattice at a trap depth of 300 lattice photon recoil energies ($E_{rec}$), loading a millimeter scale atomic sample (Fig. 1a). We then adiabatically ramp the lattice to an operational depth of 12 $E_{rec}$ and prepare atoms into a single nuclear spin. Clock interrogation proceeds by probing the ultranarrow $^1S_0$ ($g$)→ $^3P_0$ ($e$) transition with the resulting excitation fraction measured by fluorescence spectroscopy. Scattered photons are collected on a camera, enabling *in-situ* measurement with 6 μm resolution, corresponding to ~15 lattice sites (Fig. 1a).

Quantum state control has been vital to recent advances in atom-atom and atom-light coherence times in 3D OLCs and tweezer clocks[5,11,15]. Improved quantum state control is demonstrated through precision spectroscopy of the Wannier-Stark states of the OLC[29,30]. The 1D lattice oriented along gravity has the degeneracy of neighboring lattice sites lifted by the gravitational potential energy. In the limit of shallow lattice depths, this creates a set of delocalized states. By ramping the lattice depth to 6 $E_{rec}$, much lower than in traditional 1D lattice operations[7,8,10], clock spectroscopy probes this delocalization (Fig. 1d). The ability to engineer the extent of atomic wavefunctions through the adjustment of trap depth creates an opportunity to control the balance of on-site *p*-wave versus neighboring-site *s*-wave atomic interactions. We utilize this tunability by operating at a 'magic' trap depth[31], where the frequency shifts arising from on-site and off-site atomic interactions cancel, enabling a reduction of the collisional frequency shifts by more than three orders of magnitude compared our previous work[10].

Extended atomic coherence times are critical for both accuracy and precision. An aspirational milestone for clock measurement precision is the ability to coherently interrogate atomic samples up to the excited state's natural lifetime. To evaluate the limits of our clock's atomic coherence, we perform Ramsey spectroscopy to measure the decay of fringe contrast as a function of the free-evolution time. By comparing two uncorrelated regions within our atomic sample, we determine the contrast and relative phase difference between the two sub-ensembles (Fig. 2). The contrast decays exponentially with a time constant of 37 s (quality factor of $3.6 \times 10^{16}$), corresponding to an additional decoherence time of 53 s relative to the $^3P_0$ natural lifetime (118 s)[32]. This represents the longest optical atomic coherence time measured in any spectroscopy system to date.

We utilize Rabi spectroscopy in conjunction with *in-situ* imaging to microscopically probe clock transition frequencies along the entire vertically oriented atomic ensemble. With a standard interleaved probing sequence using the $|g, m_F = \pm\frac{5}{2}\rangle$ to $|e, m_F = \pm\frac{3}{2}\rangle$ transitions for minimal magnetic sensitivity, we reject the first order Zeeman shifts and vector AC Stark shifts. The *in-situ* imaging of atoms in the lattice allows measurement of unprocessed frequencies across the entire atomic sample (Fig. 1a and Methods). The dominant differential perturbations arise from atom-atom interactions (residual density shift contributions after we operate at the 'magic' trap depth) and magnetic field gradients giving rise to pixel-specific 2nd order Zeeman shifts. Using the total camera counts and $m_F$-dependent frequency splitting, we correct the density and 2nd order Zeeman shift at each pixel. These corrections result in the processed frequencies per pixel shown in Fig. 3a, with error bars representing the quadrature sum of statistical uncertainties from the center frequency, the density shift correction, and the 2nd order Zeeman shift correction. Additional systematics are described in the Methods. This approach demonstrates an efficient method for rapid and accurate evaluation of various systematic effects throughout a single atomic ensemble. Unlike traditional 1D OLCs where systematic uncertainties are quoted as global parameters, we now microscopically characterize these effects.

This new microscopic *in-situ* imaging allows determination of the gravitational redshift within a single atomic sample, probing an uncharacterized fundamental clock systematic. Two identical clocks on the surface of a planet separated by a vertical distance $h$ will differ in frequency ($\delta f$) as given by

$$\frac{\delta f}{f} = \frac{ah}{c^2} \quad \text{(Eq. 1)},$$

with $f$ the clock frequency, $c$ the speed of light, and $a$ the gravitational acceleration. The gravitational redshift at Earth's surface corresponds to a fractional frequency gradient of $-1.09 \times 10^{-19}$/mm in the coordinate system of Fig. 1a. Measurement of a vertical gradient across the atomic sample consistent with the gravitational redshift provides an exquisite verification of an individual atomic clock's frequency control.

Our intra-cloud frequency map (Fig. 3a) allows us to evaluate gradients across the atomic sample. Over 10 days we performed 14 measurements (ranging in duration from 1-17 hours) to search for the gravitational redshift across our sample. For each dataset we fit a linear slope and offset, reporting the slope in Fig. 3b. From this measurement campaign we find the weighted mean (standard error of the weighted mean) of the frequency gradient in our system to be -1.00(12)×10$^{-19}$/mm. We evaluate additional differential systematics (see Methods) and find a final frequency gradient of -9.8(2.3)×10$^{-20}$/mm, consistent with the predicted redshift.

The ability to resolve the gravitational redshift within our system suggests a level of frequency control beyond previous clock demonstrations, vital for the continued advancement of clock accuracy and precision. Previous fractional frequency comparisons[15] have reached uncertainties as low as 4.2×10$^{-19}$. Similarly, we perform a synchronous comparison between two uncorrelated regions of our atomic cloud (Fig. 4a). By binning ~100 pixels per region, we substantially reduce instability caused by quantum projection noise[33]. Analyzing the frequency difference between regions from 92 hours of data, we find a fractional frequency instability of 4.4×10$^{-18}$/$\sqrt{\tau}$ ($\tau$ is the averaging time in seconds), resulting in a fractional frequency uncertainty of 7.6×10$^{-21}$ for full measurement time, nearly two orders of magnitude lower than the previous record. From this measurement we infer a single region instability of 3.1×10$^{-18}$/$\sqrt{\tau}$. Dividing the fractional frequency difference by the spatial separation between each region's center of mass gives a frequency gradient of -1.30(18)×10$^{-19}$/mm. Correcting for additional systematics as before results in a gradient of -1.28(27)×10$^{-19}$/mm, again fully consistent with the predicted redshift.

In conclusion, we have established a new paradigm for atomic clocks. The vastly improved atomic coherence and frequency homogeneity throughout our sample allow us to resolve the gravitational redshift at the submillimeter scale, observing for the first time the frequency gradient from gravity within a single sample. We demonstrate a synchronous clock comparison between two uncorrelated regions with a fractional frequency uncertainty of 7.6×10$^{-21}$, advancing precision by nearly two orders of magnitude. These results suggest that there are no fundamental limitations to inter-clock comparisons reaching frequency uncertainties at the 10$^{-21}$ level, offering new opportunities for tests of fundamental physics.

**Fig. 1: Experimental system and quantum state control. a,** A millimeter length sample of ~100,000 $^{87}$Sr atoms are trapped in a 1D optical lattice formed within an in-vacuum cavity. The longitudinal axis of the cavity, z, is oriented along gravity. We probe atoms along the $^1S_0 \rightarrow {}^3P_0$ transition using a clock laser locked to an ultrastable crystalline silicon cavity[6,34]. **b,** Rabi spectroscopy with a 3.1 s pulse time. Open purple circles indicate data with a corresponding Rabi fit in green. **c,** Neighboring lattice sites are detuned by gravity, creating a Wannier-Stark ladder. Clock spectroscopy probes the overlap of Wannier-Stark states between lattice sites that are m sites away with Rabi frequency $\Omega_m$. **d,** Rabi spectroscopy probes Wannier-Stark state transitions, revealing wavefunction delocalization of up to 5 lattice sites. The number of lattice sites is indicated above each transition, with blue(red) denoting Wannier-Stark transitions to higher(lower) lattice sites.

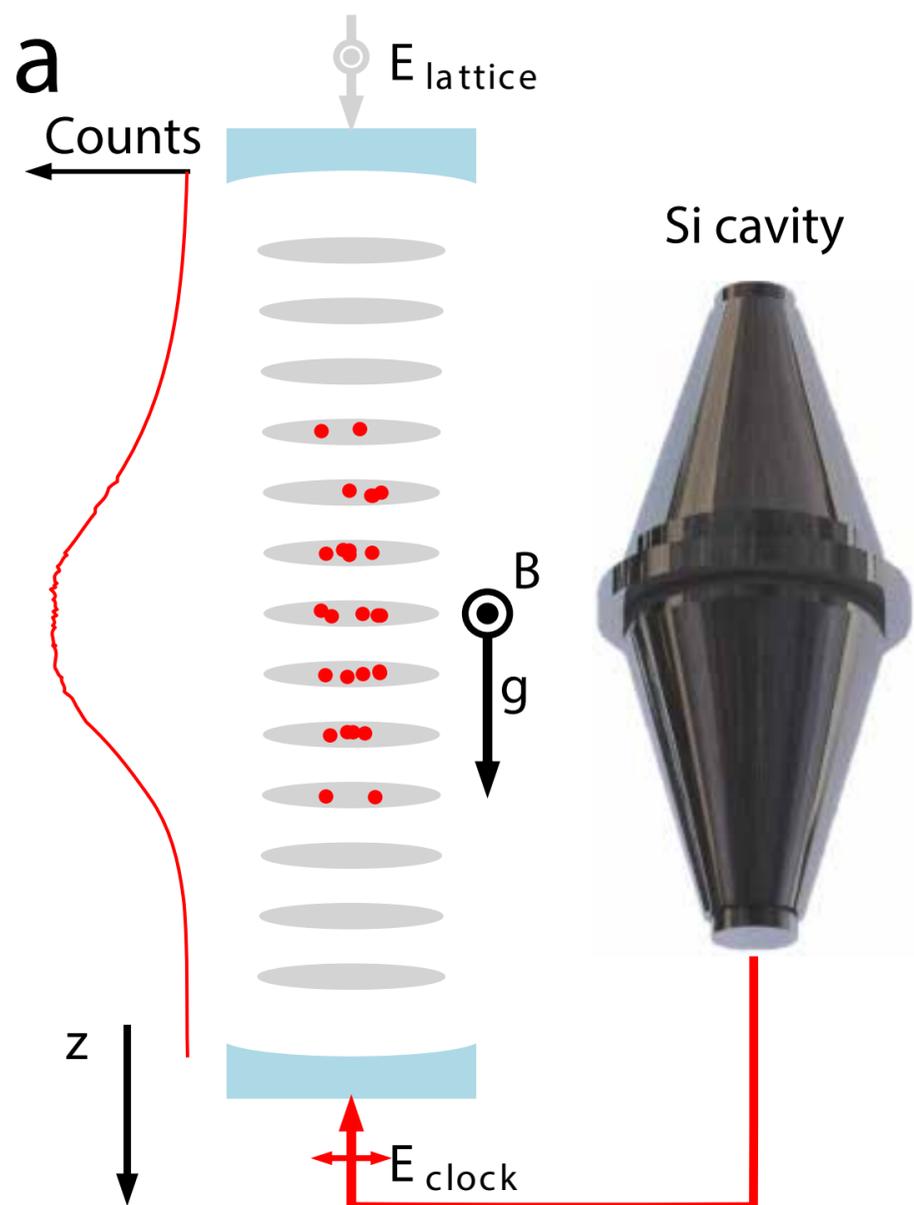
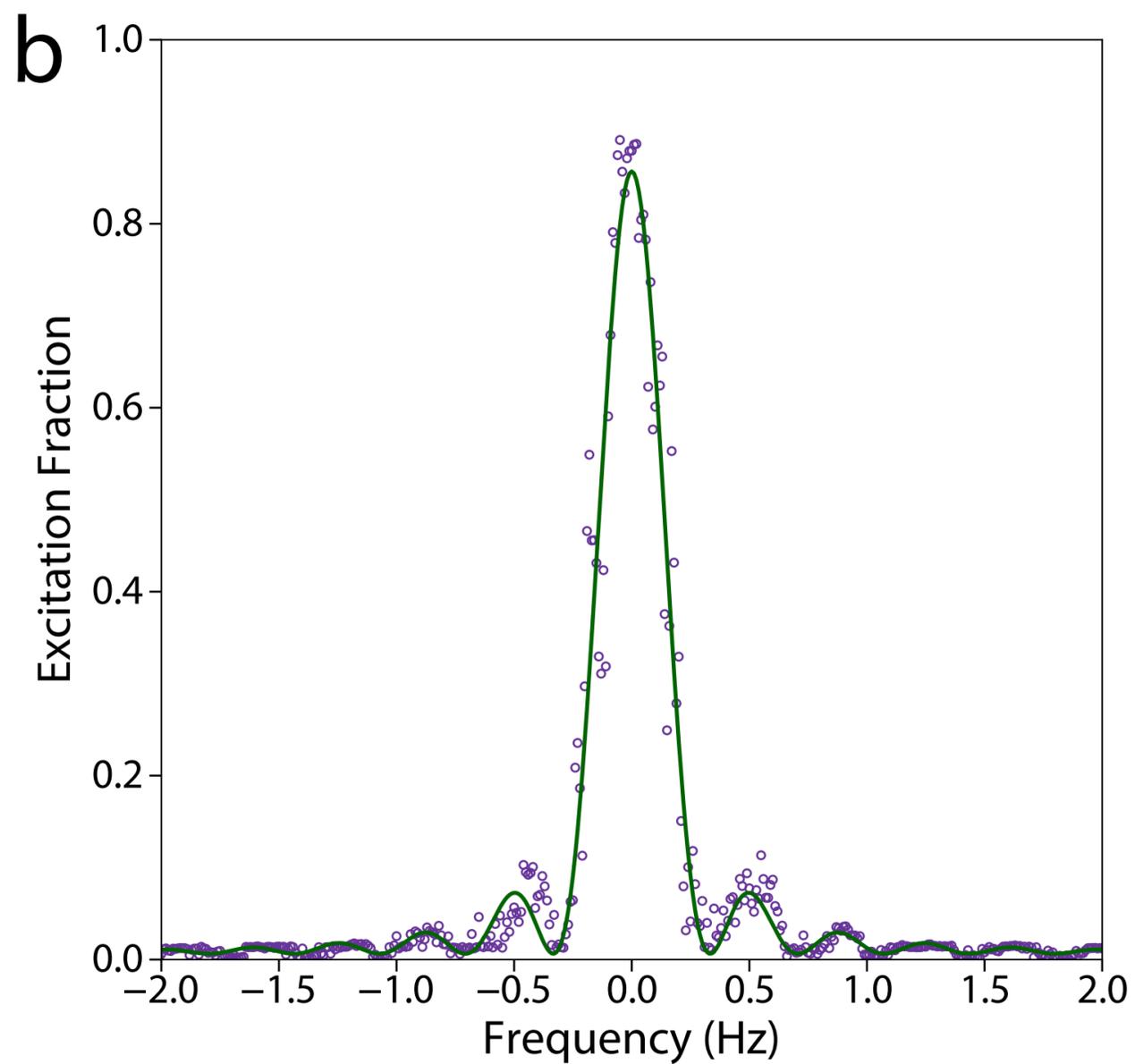
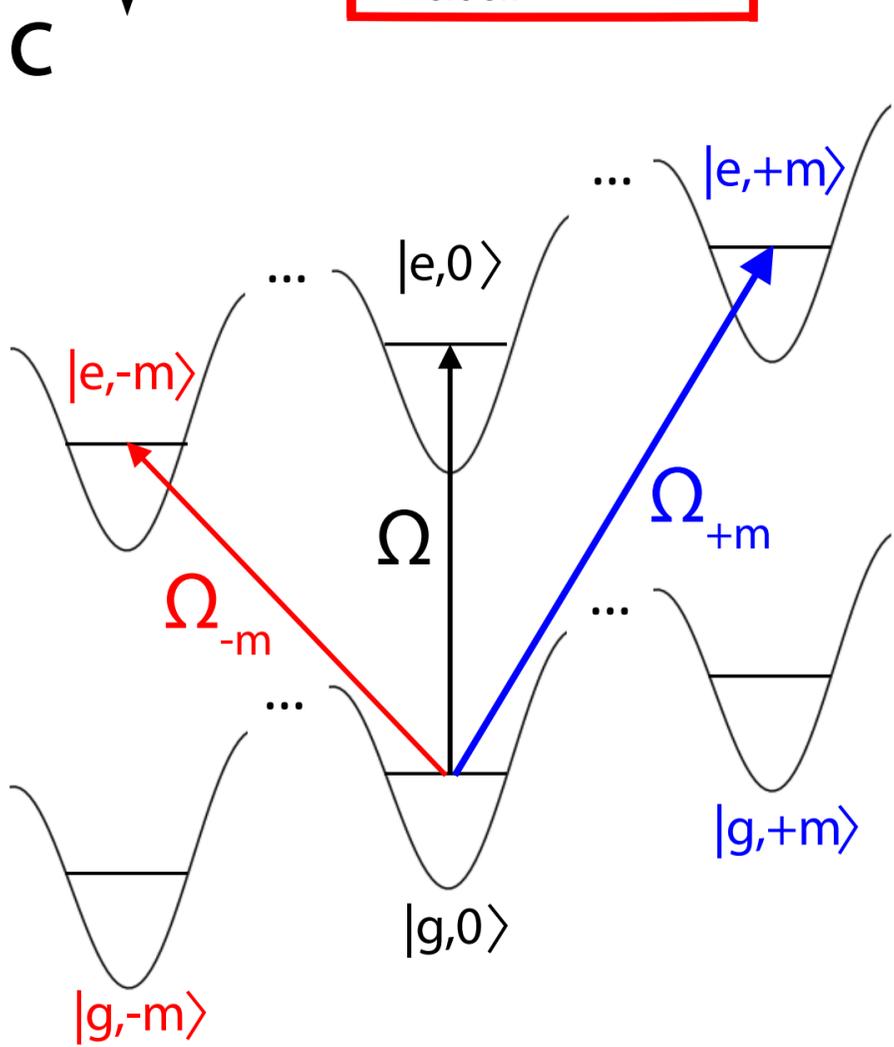
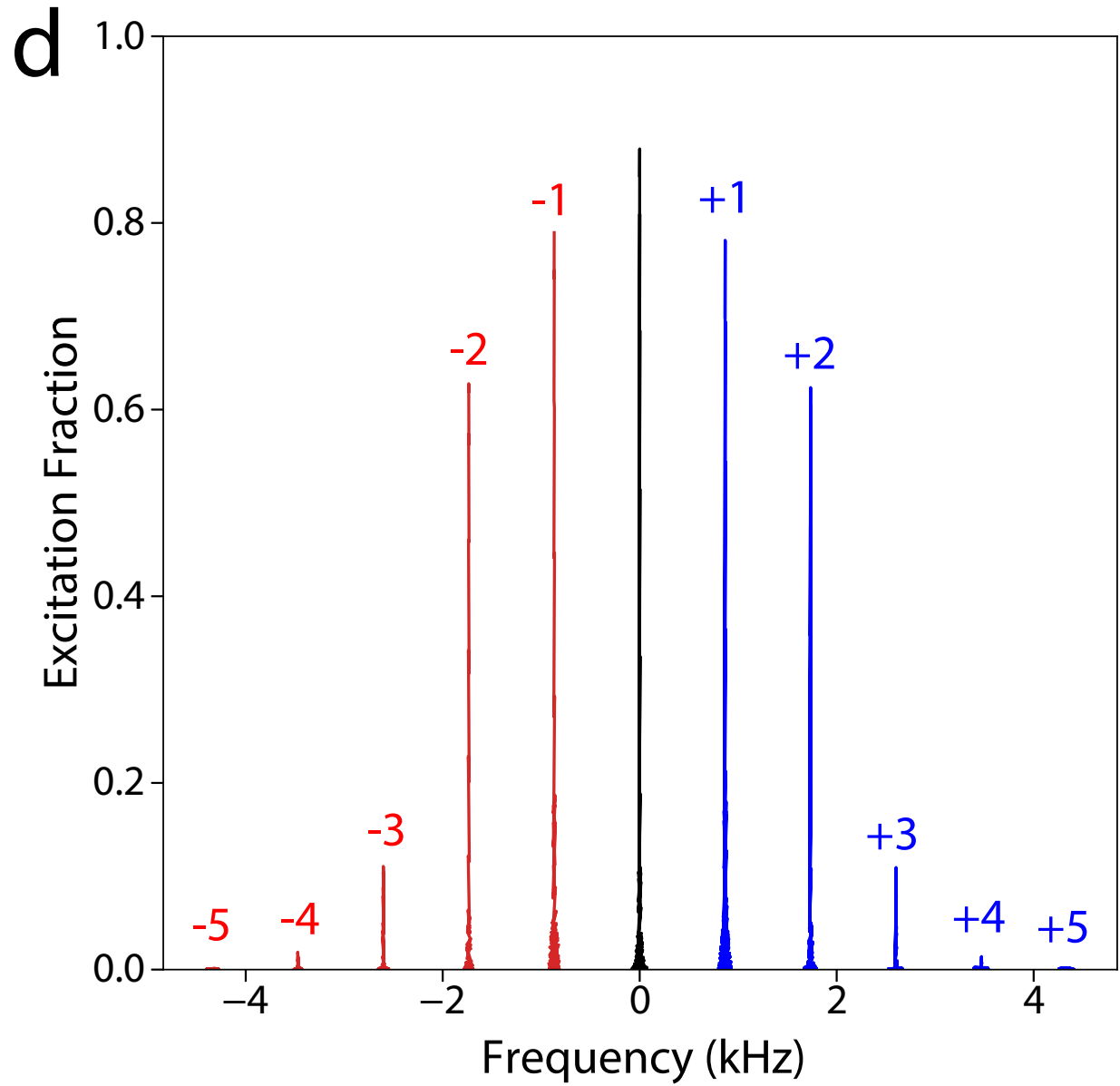

**Fig. 2: Atomic coherence.** We use Ramsey spectroscopy with a randomly sampled phase for the second pulse to determine the coherence time of our system[11]. **a,** We measure the excitation fraction across the cloud, shown in purple for a single measurement, and calculate the average excitation fractions in regions $p_1$ and $p_2$, separated by 2 pixels. **b,** Parametric plots of the excitation fraction of $p_1$ versus $p_2$ in purple for 6 s, 30 s and 50 s dark time demonstrate a phase shift between the two regions and contrast decay. Using a maximum likelihood estimator, we extract the phase and contrast for each dark time with the fit, shown in green. **c,** Contrast decay as a function of time in green is fit with an exponential decay in gold, giving an atomic coherence decay time of 36.5(0.7) s and a corresponding quality factor of $3.6 \times 10^{16}$. After accounting for the finite radiative decay contribution, we infer an additional decoherence time constant of 52.8(1.5) s.

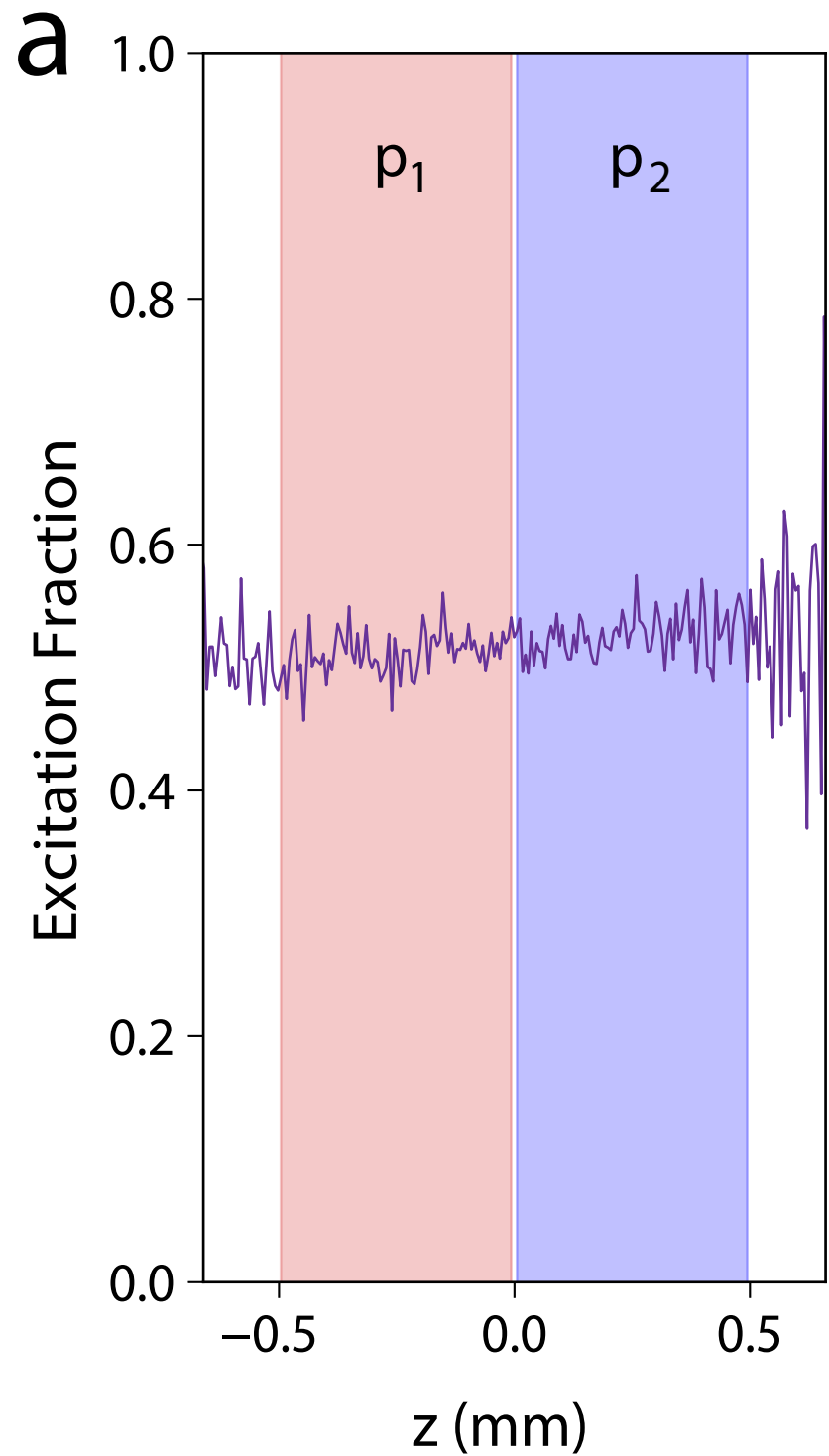
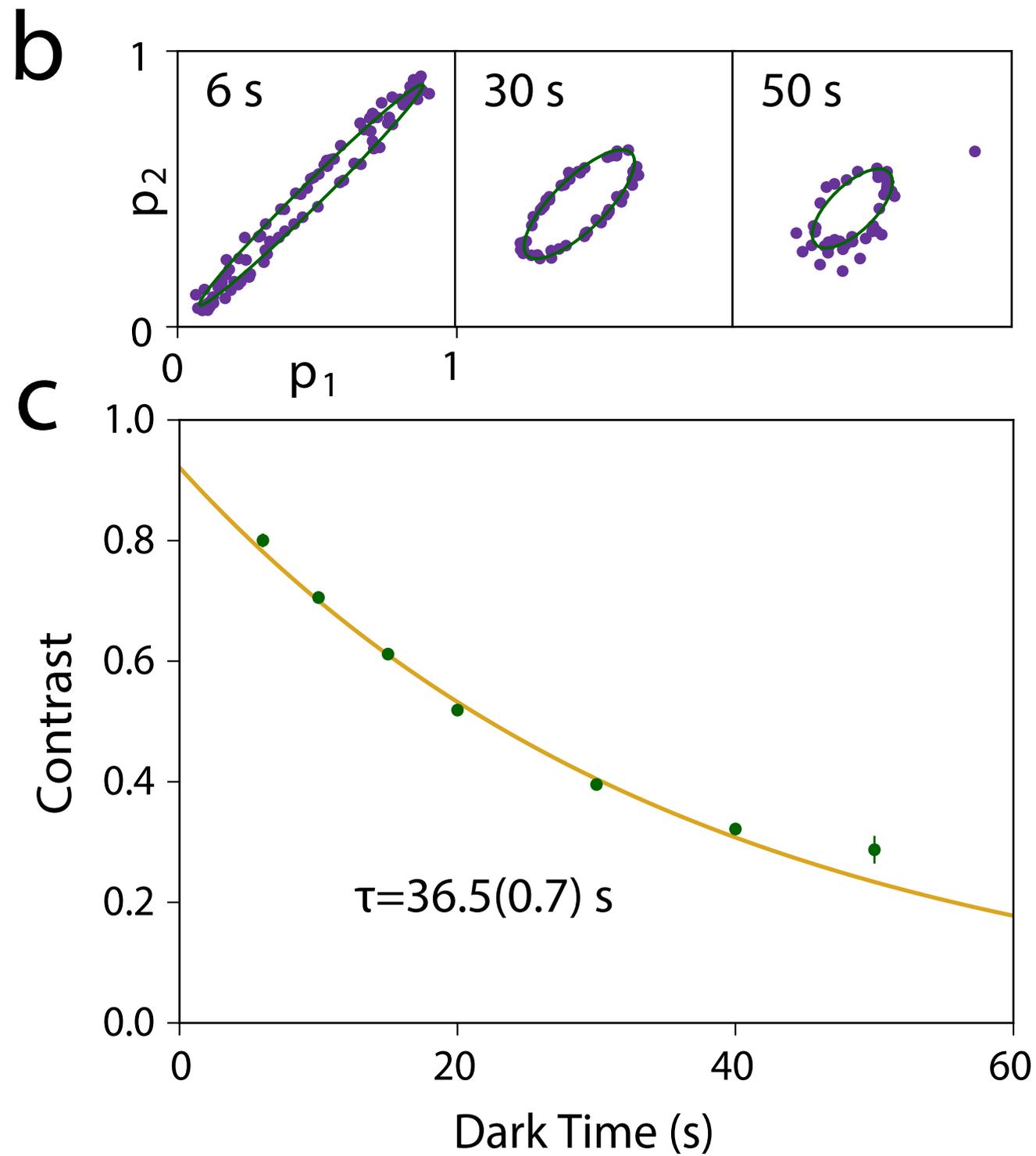

**Figure 3: Evaluating frequency gradients. a,** For each measurement we construct a microscopic frequency map across the sample, with raw frequencies shown in green. The $2^{nd}$ order Zeeman correction is shown as a dashed gold line. Processed frequencies shown in purple include both density shift corrections and $2^{nd}$ order Zeeman corrections, with uncertainties arising from the quadrature sum of statistical, density shift correction, and $2^{nd}$ order Zeeman correction uncertainties. To this we fit a linear function, shown in black. **b,** Over the course of 10 days, we completed 14 measurements. For each measurement, we create a corrected frequency map and fit a linear slope as in **a**. This slope is plotted for each measurement, as well as a weighted mean (black) with associated statistical uncertainty (dashed black). The expected gravitational gradient is shown in dashed red. All data is taken with Rabi spectroscopy using a 3.1 s π-pulse time except for 08/13 which used a 3.0 s pulse time.

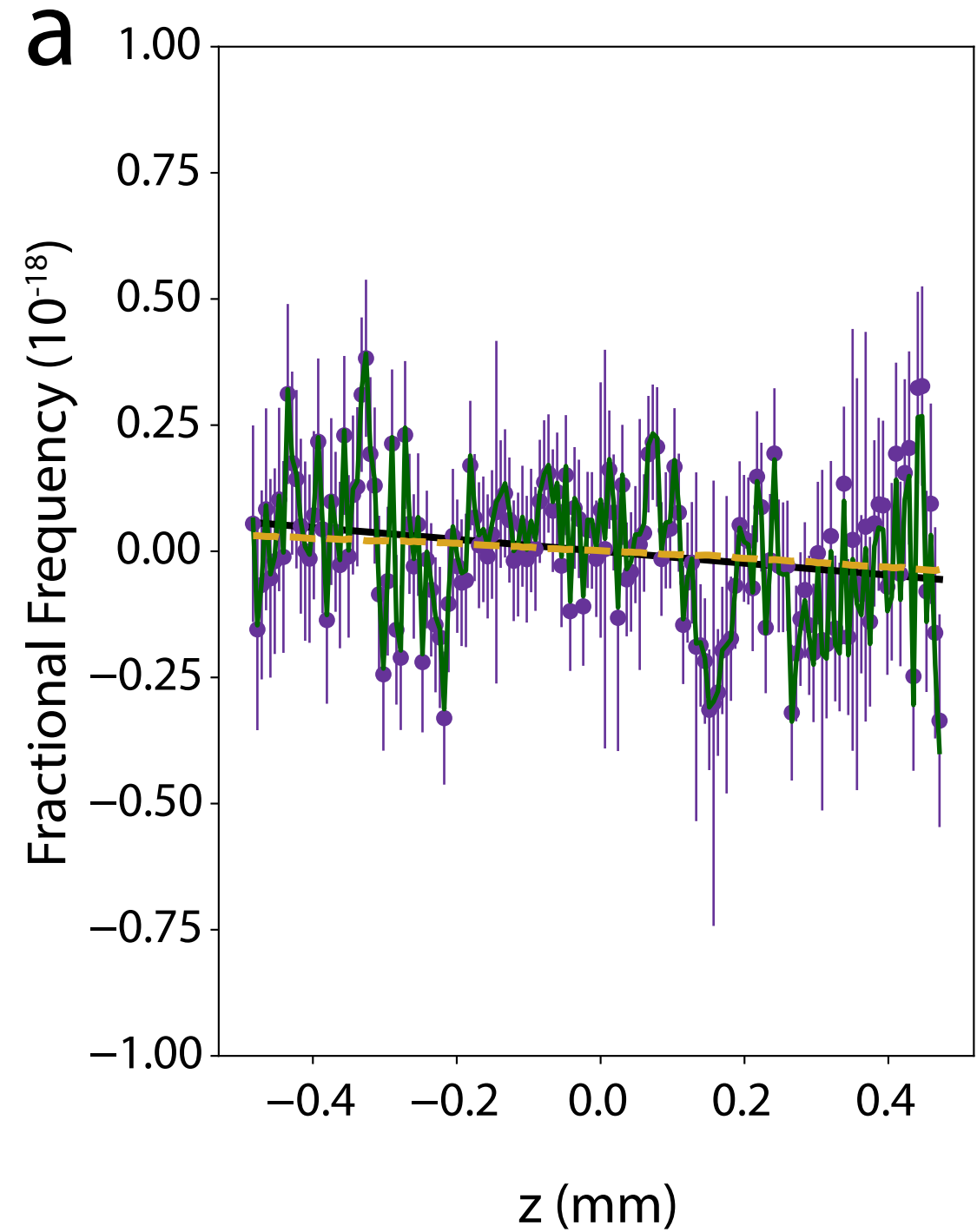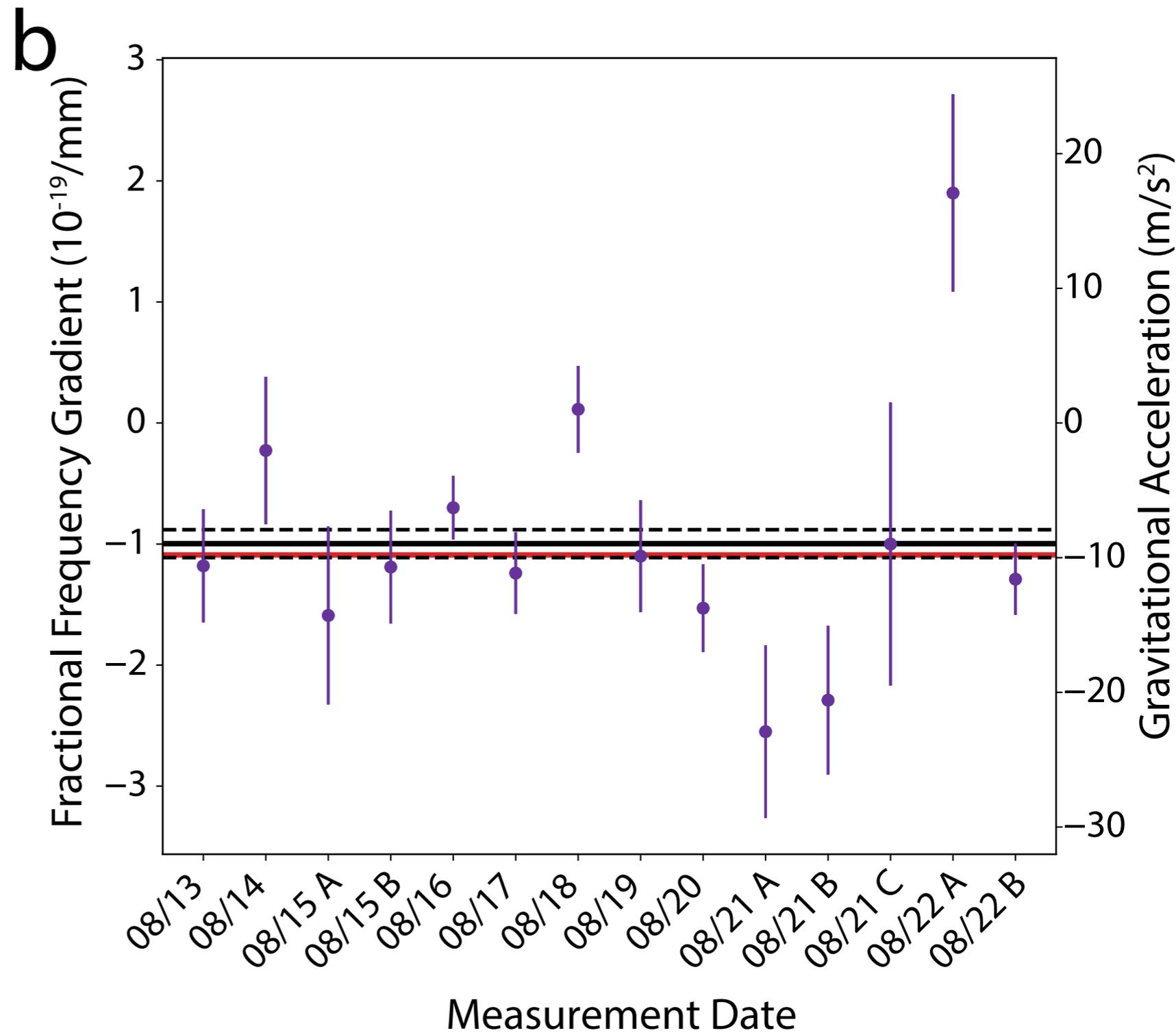

**Figure 4: *In-situ* synchronous clock comparison. a,** The cloud is separated as in Fig. 2a. The gravitational redshift leads to the higher clock(blue) ticking faster than the lower one(red). The length scale is in millimeters. **b,** Allan deviation of the frequency difference between the two regions in **a** over 92 hours. Purple points show differential stability fit by the green line. The single atomic region instability is shown in gold.

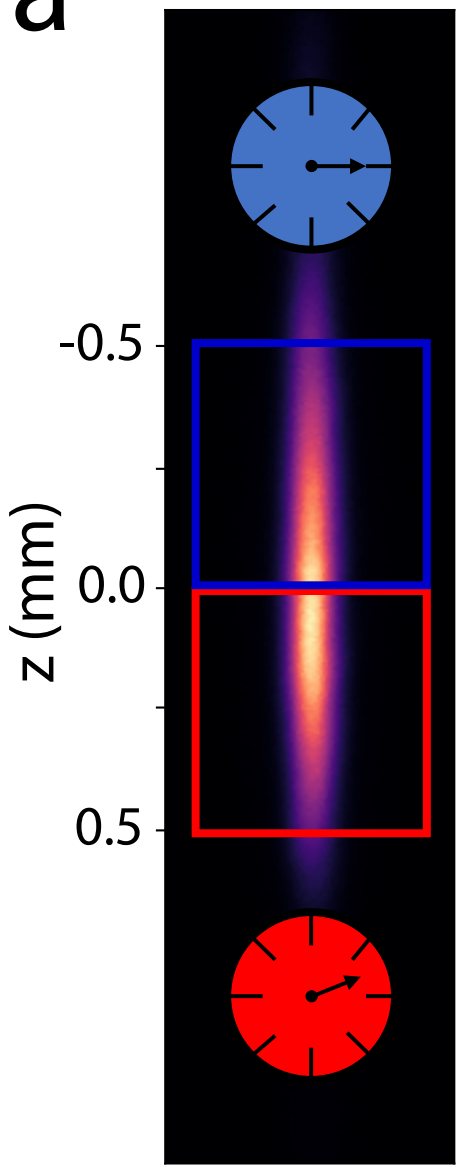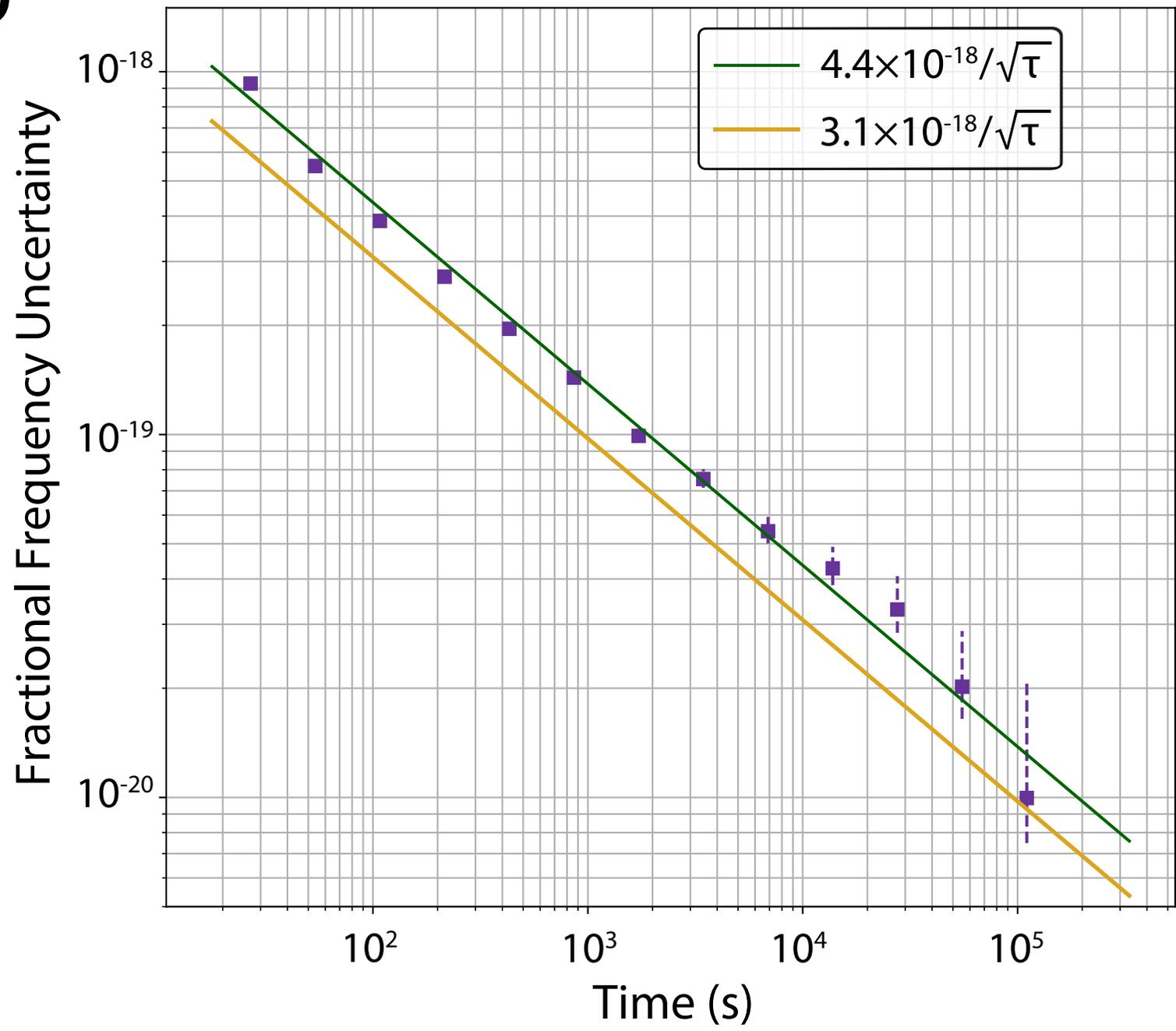

# Methods

## In-Vacuum Cavity

Central to our system is an in-vacuum lattice buildup cavity oriented along gravity (Fig. 1a). Two mirrors with radius of curvature of 1 m are separated by ~15 cm, achieving a mode waist of 260 μm. Our cavity has a finesse at the lattice wavelength (813 nm) of ~1300 and a power buildup factor of ~700. This enables lattice depths in excess of 500 $E_{rec}$ (lattice photon recoil energy) using a diode-based laser system. The dimensional stability of the cavity combined with the simplified diode laser system enables robust operation compared with our previous Ti:Sapphire retro-reflected design[10]. The cavity mirrors are anti-reflection coated at the clock wavelength of 698 nm.

One cavity mirror is mounted to a piezo for length stabilization while the other is rigidly mounted for phase reference for the clock laser. Grounded copper shields surrounding each mirror prevent DC Stark induced shifts due to charge buildup on the mirrors and piezo[35,36]. This shielding is verified through evaluation of the DC Stark shift systematic.

## Atomic sample preparation

[87]Sr atoms are cooled and loaded into a 300 $E_{rec}$ optical lattice using standard two stage magneto-optical trapping techniques[10]. Once trapped, atoms are simultaneously nuclear spin polarized, axially sideband cooled, and radially doppler cooled into a single nuclear spin state at temperatures of 800 nK. The lattice is then adiabatically ramped to the operational trap depth of 12 $E_{rec}$, where a series of pulses addressing the clock transition prepares atoms into $|g, m_F =$

$\pm\frac{5}{2}\rangle$. Clock spectroscopy is performed by interrogating the $|g, m_F = \pm\frac{5}{2}\rangle$ to $|e, m_F = \pm\frac{3}{2}\rangle$ transition, the most magnetically insensitive $^{87}$Sr clock transition[6].

## Imaging

The clock excitation fraction is read out using standard fluorescence spectroscopy techniques[6,10,37]. Photons are collected on both a photo-multiplier tube for global readout and electron multiplying charge coupled device camera for an *in-situ* readout of clock frequency. Camera readout is performed in full vertical binning mode, averaging the radial dimension of the lattice. This provides 1D *in-situ* imaging for all synchronous evaluations.

We use a 25 μs fluorescence probe with an intensity of $I/I_{sat}$ ~ 20 ($I_{sat}$ being the saturation intensity), ensuring uniform scattering across the atomic sample. Before imaging, the optical lattice is ramped back to 300 $E_{rec}$ to decrease imaging aberration resulting from the extended radial dimension at 12 $E_{rec}$.

## Analysis

Standard clock lock techniques and analysis are used[6,11,33], with differences in excitation fraction converted to frequency differences using Rabi lineshapes. Each dataset is composed of a series of clock locks, tracking the center of mass frequency of the atomic sample. A clock lock is four measurements probing alternating sides of the Rabi lineshape for opposite nuclear spin transitions. To avoid ambiguous frequency corrections due to the Rabi lineshape, we remove data with excitation fractions above (below) .903×*C* (.116×*C*), where C is the Rabi contrast. From each clock lock, a pixel specific center frequency $f_i$ and frequency splitting between opposite $m_F$ states $\Delta_i$ are calculated, creating an *in-situ* frequency map of the 1D atomic sample. This allows rejection of vector shifts on a pixel-by-pixel basis and probes the magnetic field induced splitting of $m_F$ transitions. The atom weighted mean frequency is subtracted from every lock cycle to reject common mode laser noise.

For each dataset we approximate the atomic profile with a Gaussian fit, identifying a center pixel and associated Gaussian width ($\sigma$). All analysis is performed within the central region of $\pm 1.5\sigma$ which demonstrates the lowest frequency instability. Identifying a center pixel for data processing ensures rejection of any day-to-day drift in the position of the cloud due to varying magnetic fields modifying MOT operation on the narrow line transition. The density shift coefficient (see Density Shift section) is derived from the average center frequency per pixel. Using this coefficient, we correct $f_i$ and $\Delta_i$ for the density shift. $2^{nd}$ order Zeeman corrections using these updated frequencies are then applied.

Gradient analysis is based on the processed center frequencies per pixel. A linear fit to the frequencies as a function of pixel is performed using least squares, with uncertainty per pixel arising from the quadrature sum of statistical frequency uncertainty, statistical $2^{nd}$ order Zeeman uncertainty, and density shift correction uncertainty.

For the two-clock comparison (Fig. 4b), all data from 8/14-8/22 was taken with the same duty cycle and π-pulse time (3.1 s). Data was processed relative to a fit center pixel as discussed and concatenated. Two equal regions extend from the center of the sample to a width of $\pm 1.5\sigma$, with two empty pixels between regions to ensure uncorrelated samples. Each region is processed for the atom weighted mean frequency, enabling a synchronous frequency comparison between two independent clocks.

## Atomic Coherence

We use a Ramsey sequence to measure the atomic coherence. We prepare a sample in the $|g, m_F = +\frac{5}{2}\rangle$ state and apply a π/2 pulse along the $|g, m_F = +\frac{5}{2}\rangle$ to $|e, m_F = +\frac{3}{2}\rangle$ transition. After waiting for a variable dark time, we apply a second π/2 pulse with a random phase relative to the first. We then measure the excitation fraction.

Two regions, $p_1$ and $p_2$, are identified using the same technique as in the synchronous instability measurement. For each experimental sequence, we find the average excitation fraction in $p_1$ and $p_2$. A mean frequency shift across the sample primarily due to a magnetic field gradient creates a differential phase as a function of time between $p_1$ and $p_2$. We create a parametric plot of the average excitation in $p_1$ and $p_2$ for each dark time and use a maximum likelihood estimator to fit an ellipse to each dataset, calculating phase and contrast[15,38]. To estimate uncertainty in the contrast for each dark time a bootstrapping technique is used[11]. Fitting the contrast as a function of dark time with a single exponential returns an effective atomic coherence time.

## Systematics

### Imaging

We calibrate our pixel size using standard time of flight methods: we observe an atomic sample in freefall for varying times to determine an effective pixel size along the direction of gravity. Immediately after our 10-day data campaign we measured our effective pixel size to be 6.04 µm. Due to thermal drift of our system, the pixel size can vary by up to 0.5 µm /pixel over months which we take as the calibration uncertainty.

Spatial correlations may limit imaging resolution. We measure these correlations by placing atoms into a superposition of clock states. Any measured spatial correlation is due to the imaging procedure. In our system we find no correlations between neighboring pixels[11]. The optical resolution of our imaging lens is specified at 2 µm.

Lattice tilt from gravity will modify the measured gradient. We find the lattice tilt in the imaging plane to be 0.11(0.06) degrees, providing an uncertainty orders of magnitude smaller than the pixel size uncertainty. We are insensitive to lattice tilt out of the imaging plane.

**Zeeman Shifts**

First order Zeeman shifts are rejected by probing opposite nuclear spin states[6]. The 2nd order Zeeman shift is given by

$$\Delta v_{B,2} = \xi (\Delta v_{B,1})^2,$$

where $\Delta v_{B,1}$ is the splitting between opposite spin states and $\xi$ the corresponding 2nd order Zeeman shift coefficient. For stretched spin state operation ($m_F = \pm \frac{9}{2}$), $\xi$=-2.456(3)×10$^{-7}$ Hz. Using known atomic coefficients[39] we find the 2nd order Zeeman coefficient for the $|g, m_F = \pm \frac{5}{2}\rangle$ to $|e, m_F = \pm \frac{3}{2}\rangle$ transition to be $\xi_{op}$=-1.23(8)×10$^{-4}$ Hz, with the uncertainty arising from limited knowledge of atomic coefficients. We include an error of 4×10$^{-21}$/mm in Table 1 to account for this atomic uncertainty.

**DC Stark**

Electric fields perturb the clock frequency via the DC Stark effect. We evaluate this shift by using in-vacuum quadrant electrodes to apply bias electric fields in all three dimensions. We find a DC Stark gradient of 3(2)×10$^{-21}$/mm.

**Black Body Radiation Shift**

The dominant frequency perturbation to room temperature neutral atom clocks is black body radiation (BBR). Similar to our previous work[10], we homogenize this shift by carefully controlling the thermal surroundings of our vacuum chamber. Attached to the vacuum chamber are additional temperature control loops, with each vacuum viewport having a dedicated temperature control system. This ensures our dominant BBR contribution – high emissivity glass viewports – are all the same temperature to within 100 mK.

To bound possible BBR gradients, we introduce a 1 K gradient between the top and bottom of the chamber along the cavity axis by raising either the top or bottom viewports by 1 K. We compare these two cases and find no statistically significant changes in the frequency gradient across the entire sample. Accounting for uncertainty in linear frequency fits for each case, we estimate an uncertainty of 3×10$^{-21}$/mm. This finding is supported with a basic thermal model of the vacuum chamber.

**Density Shift**

Atomic interactions during Rabi spectroscopy lead to clock frequency shifts as a function of atomic density[40]. For each gradient measurement, we evaluate the density shift coefficient $\chi_{dens}$ by fitting the average frequency $f$ per pixel versus average camera counts per pixel $N$ to an equation of the form

$$f(N) = \chi_{dens} N + B.$$

Here B is an arbitrary offset. Once $\chi_{dens}$ is known, we remove the density shift at each pixel.

Residual density shift corrections may lead to error in our linear gradient. To bound this effect, we compare the density shift coefficient and gradient from our data run with a separate dataset at 8 E$_{rec}$. With the trap depth at 8 E$_{rec}$ we found a linear gradient of s=-1.08×10$^{-18}$/mm and a density shift coefficient of $\chi_8$=-1.39×10⁻⁶ Hz/count. During our data run we had an average density shift coefficient of $\chi_{op}$=-2.43×10⁻⁸ Hz/count. We bound the uncertainty in our gradient from density shift as $\sigma_{den,unc} = s \times \frac{\chi_{op}}{\chi_8}$=1.7×10⁻²⁰/mm.

**Lattice Light Shifts**

Lattice light shifts arise from differential AC Stark shifts between the ground and excited clock states. An approximate microscopic model of the lattice light shift ($\nu_{LS}$) in our system is given by[41]

$$h\nu_{LS}(u,\delta_L) \approx \left(\frac{\delta\Delta\alpha^{E1}}{\delta\nu}\delta_L - \Delta\alpha_{QM}\right)\frac{u^{\frac{1}{2}}}{2} - \left[\frac{\delta\alpha^{E1}}{\delta\nu}\delta_L\right]u,$$

where $u$ is the trap depth in units of E$_{rec}$, $\Delta\alpha^{E1}$ the differential electric dipole polarizability, $\Delta\alpha^{QM}$ the differential multi-polarizability, and $\delta_L = (\nu_L - \nu^{E1})$ the detuning between lattice frequency $\nu_L$ and effective magic frequency $\nu^{E1}$. Our model has no dependence on the longitudinal vibrational quanta since we are in the ground vibrational band. We neglect higher order corrections from hyperpolarizability due to our operation at depths <60 E$_{rec}$. At our temperatures thermal averaging of the trap depth is a higher order correction (<5%) that is also neglected.

We model the linear differential lattice light shift across the atomic cloud as

$$\frac{\delta h\nu_{LS}(u,\delta_L)}{\delta z} \approx \left[\frac{\left(\frac{\delta\alpha^{E1}}{\delta\nu}\delta_L - \alpha_{QM}\right)}{4u^{1/2}} - \frac{\delta\alpha^{E1}}{\delta\nu}\delta_L\right]\frac{\delta u}{\delta z},$$

where z is the coordinate corresponding to the axis of the cavity along gravity. To evaluate our differential lattice light shift at our operational depth we need $\delta_L$ and $\frac{\delta u}{\delta z}$. We modulate our lattice between two trap depths ($u_1$=14 $E_{rec}$, $u_2$=56 $E_{rec}$) and find our detuning from scalar magic frequency to be $\delta_L$=7.4(0.6) MHz. To evaluate $\frac{\delta u}{\delta z}$ at our operational depth ($u_{op}$) we measure the linear gradient across the atomic cloud at $\delta_L$ + 250 MHz and $\delta_L$- 250 MHz, the difference given by

$$\frac{\delta h \nu_{LS}(u_{op}, \delta_{L+250\ MHz})}{\delta z} - \frac{\delta h \nu_{LS}(u_{op}, \delta_{L-250\ MHz})}{\delta z} \approx \left[\frac{1}{4u_{op}^{\frac{1}{2}}} - 1\right] \frac{\delta \alpha^{E1}}{\delta \nu} \delta_{500} \frac{\delta u_{op}}{\delta z},$$

where $\delta_{500}$= 500 MHz. We find $\frac{\delta u_{op}}{\delta z}$ =0.0383/mm, which when combined with $\delta_L$=7.4 MHz, gives us a fractional frequency gradient of -5×10$^{-21}$/mm. Accounting for error in our lattice detuning and linear gradient gives us an uncertainty of 1×10$^{-21}$/mm.

**Other Systematics**

For a 3.1 s π-pulse the probe AC Stark shift[7] is -3(2)×10$^{-21}$. A frequency scan of the $|g, m_F = -\frac{5}{2}\rangle$ to $|e, m_F = -\frac{3}{2}\rangle$ transition limits the variation of excitation fraction across the atomic sample to 1% or below, bounding any possible probe AC Stark gradient across the sample to <1×10$^{-22}$.

**Known Redshift**

The gravitational acceleration (rounding to 4 digits) within our lab was evaluated by a USGS survey[42] to be a=-9.796 m/s².

**Systematic Budget**

| Systematic | Slope (10$^{-20}$/mm) | Uncertainty (10$^{-20}$/mm) |
|---|---|---|
| Statistical (Fig 3) | -10.0 | 1.2 |
| BBR | 0 | 0.3 |

| Density | - | 1.7 |
|---|---|---|
| Lattice light shift | -0.5 | 0.1 |
| DC Stark | 0.3 | 0.2 |
| Pixel Calibration | 0 | .8 |
| 2nd Order Zeeman | - | .4 |
| Other | 0 | <.1 |
| Corrected Gradient | -9.8 | 2.3 |
| Known Redshift | -10.9 | <.1 |

**Table 1: Gradient Systematic Budget.** Fractional frequency gradients and corresponding uncertainties. Fractional frequencies denoted with '-' are corrected on a pixel-by-pixel basis during initial data processing. The corrected gradient has known systematics removed with uncertainty given by the quadrature sum of all correction uncertainties.

**Acknowledgements**

We acknowledge funding support from Defense Advanced Research Projects Agency, National Science Foundation QLCI OMA-2016244, DOE Quantum System Accelerator, NIST, NSF Phys-1734006, and Air Force Office for Scientific Research. We are grateful for theory insight from Anjun Chu, Peiru He, and Ana María Rey. We acknowledge stimulating discussion and technical contributions from John Zaris, James Uhrich, Ross Hutson, Christian Sanner, William Milner, Lindsay Sonderhouse, Lingfeng Yan, Maya Miklos, Yee Ming Tso, and Shimon Kolkowitz. We thank James Thompson, Cindy Regal, John Hall, and Serge Haroche for careful reading of the manuscript.

**Authors' note:** While performing the work described here, we became aware of complementary work where high measurement precision was achieved for simultaneous differential clock comparisons between multiple atomic ensembles in vertical 1D lattices separated by centimeter scale distances using a hertz-linewidth clock laser[43].

## Author Contributions

All authors contributed to carrying out the experiments, interpreting the results, and writing the manuscript.

*tobias.bothwell@colorado.edu, ye@jila.colorado.edu

## Competing interests

The authors declare no competing interests.

## References

1. Einstein, A. Grundgedanken der allgemeinen Relativitätstheorie und Anwendung dieser Theorie in der Astronomie. *Preuss. Akad. der Wissenschaften, Sitzungsberichte* **315**, (1915).
2. Chou, C. W., Hume, D. B., Rosenband, T. & Wineland, D. J. Optical clocks and relativity. *Science* **329**, 1630–1633 (2010).
3. Herrmann, S. *et al.* Test of the Gravitational Redshift with Galileo Satellites in an Eccentric Orbit. *Phys. Rev. Lett.* **121**, 231102 (2018).
4. Delva, P. *et al.* Gravitational Redshift Test Using Eccentric Galileo Satellites. *Phys. Rev. Lett.* **121**, 231101 (2018).
5. Campbell, S. L. *et al.* A Fermi-degenerate three-dimensional optical lattice clock. *Science* **358**, 90–94 (2017).
6. Oelker, E. *et al.* Demonstration of $4.8\times10^{-17}$ stability at 1 s for two independent optical clocks. *Nat. Photonics* **13**, 714–719 (2019).
7. Nicholson, T. *et al.* Systematic evaluation of an atomic clock at $2\times10^{-18}$ total uncertainty. *Nat. Commun.* **6**, (2015).

8. McGrew, W. F. *et al.* Atomic clock performance enabling geodesy below the centimetre level. *Nature* **564**, 87–90 (2018).

9. Brewer, S. M. *et al.* $^{27}$Al$^+$ Quantum-Logic Clock with a Systematic Uncertainty below $10^{-18}$. *Phys. Rev. Lett.* **123**, 33201 (2019).

10. Bothwell, T. *et al.* JILA SrI optical lattice clock with uncertainty of $2.0\times10^{-18}$. *Metrologia* **56**, (2019).

11. Marti, G. E. *et al.* Imaging Optical Frequencies with 100 µHz Precision and 1.1 µm Resolution. *Phys. Rev. Lett.* **120**, (2018).

12. Pedrozo-peñafiel, E. *et al.* Entanglement on an optical atomic-clock transition. *Nature* **588**, (2020).

13. Kaubruegger, R. *et al.* Variational Spin-Squeezing Algorithms on Programmable Quantum Sensors. *Phys. Rev. Lett.* **123**, 260505 (2019).

14. Kómár, P. *et al.* Quantum Network of Atom Clocks: A Possible Implementation with Neutral Atoms. *Phys. Rev. Lett.* **117**, 060506 (2016).

15. Young, A. W. *et al.* Half-minute-scale atomic coherence and high relative stability in a tweezer clock. *Nature* **588**, 408–413 (2020).

16. Safronova, M. S. *et al.* Search for new physics with atoms and molecules. *Rev. Mod. Phys.* **90**, 25008 (2018).

17. Sanner, C. *et al.* Optical clock comparison for Lorentz symmetry testing. *Nature* **567**, 204–208 (2019).

18. Kennedy, C. J. *et al.* Precision Metrology Meets Cosmology: Improved Constraints on Ultralight Dark Matter from Atom-Cavity Frequency Comparisons. *Phys. Rev. Lett.* **125**, 1–10 (2020).

19. Atomic, B. & Optical, C. Frequency ratio measurements at 18-digit accuracy using an optical clock network. **591**, (2021).

20. Kolkowitz, S. *et al.* Gravitational wave detection with optical lattice atomic clocks. *Phys. Rev. D* **94**, 124043 (2016).

21. Hafele, J. C. & Keating, R. E. Around-the-World Atomic Clocks: *Science* **177**, 166 (1972).

22. Takamoto, M. *et al.* Test of general relativity by a pair of transportable optical lattice clocks. *Nat. Photonics* **14**, 411–415 (2020).

23. Laurent, P., Massonnet, D., Cacciapuoti, L. & Salomon, C. The ACES/PHARAO space mission. *Comptes Rendus Phys.* **16**, 540–552 (2015).

24. Tino, G. M. *et al.* SAGE : A proposal for a space atomic gravity explorer. *Eur. Phys. J. D* **73,** 228 (2019).

25. Grotti, J. *et al.* Geodesy and metrology with a transportable optical clock. *Nat. Phys.* **14**, 437–441 (2018).

26. Pavlis, N. K. & Weiss, M. A. A re-evaluation of the relativistic redshift on frequency standards at NIST, Boulder, Colorado, USA. *Metrologia* **54**, 535–548 (2017).

27. Kolkowitz, S. *et al.* Spin-orbit-coupled fermions in an optical lattice clock. *Nature* **542**, 66–70 (2017).

28. Bromley, S. L. *et al.* Dynamics of interacting fermions under spin-orbit coupling in an optical lattice clock. *Nat. Phys.* **14**, 399–404 (2018).

29. Wilkinson, S. R., Bharucha, C. F., Madison, K. W., Niu, Q. & Raizen, M. G. Observation of atomic wannier-stark ladders in an accelerating optical potential. *Phys. Rev. Lett.* **76**, 4512–4515 (1996).

30. Lemonde, P. & Wolf, P. Optical lattice clock with atoms confined in a shallow trap. *Phys. Rev. A - At. Mol. Opt. Phys.* **72**, 1–8 (2005).

31. Aeppli, A. *et al.*, In Preparation (2021).

32. Muniz, J. A., Young, D. J., Cline, J. R. K. & Thompson, J. K. Cavity-QED measurements of the $^{87}$Sr millihertz optical clock transition and determination of its natural linewidth. *Phys. Rev. Res.* **3**, 023152 (2021).

33. Ludlow, A. D., Boyd, M. M., Ye, J., Peik, E. & Schmidt, P. O. Optical atomic clocks. *Rev. Mod. Phys.* **87**, 637–701 (2015).

34. Matei, D. G. *et al.* 1.5 μm Lasers with Sub-10 mHz Linewidth. *Phys. Rev. Lett.* **118**, 263202 (2017).

35. Lemonde, P., Brusch, A., Targat, R. Le, Baillard, X. & Fouche, M. Hyperpolarizability Effects in a Sr Optical Lattice Clock. *Phys. Rev. Lett.* **96**, 103003 (2006).

36. Lodewyck, J., Zawada, M., Lorini, L., Gurov, M. & Lemonde, P. Observation and Cancellation of a Perturbing dc Stark Shift in Strontium Optical Lattice Clocks. **59**, 411–415 (2012).

37. Leibfried, D., Blatt, R., Monroe, C. & Wineland, D. Quantum dynamics of single trapped ions. *Rev. Mod. Phys.* **75**, 281–324 (2003).

38. Marti, G. E. *et al.* Imaging Optical Frequencies with 100 μhz Precision and 1.1 μm Resolution. *Phys. Rev. Lett.* **120**, 103201 (2018).

39. Boyd, M. M. *et al.* Nuclear spin effects in optical lattice clocks. *Phys. Rev. A* **76**, 022510 (2007).

40. Martin, M. J. *et al.* A quantum many-body spin system in an optical lattice clock. *Science* **341**, 632–636 (2013).

41. Ushijima, I. *et al.* Operational Magic Intensity for Sr Optical Lattice Clocks. *Phys. Rev. Lett.* **121**, 263202 (2018).

42. van Westrum, D. NOAA Technical Memorandum NOS NGS-77. (2019).

43. Xin Zheng Jonathan Dolde, V. L. B. M. H. L. & Kolkowitz, S. High precision differential clock comparisons with a multiplexed optical lattice clock (submitted 2021).